\documentclass[
 amsmath,amssymb,
floatfix,
aip,rsi,reprint,graphicx]{revtex4-2}

\usepackage{graphicx}% Include figure files
\usepackage{dcolumn}% Align table columns on decimal point
\usepackage{bm}% bold math
\usepackage{amsmath}
\usepackage{physics}
\usepackage{svg}
\usepackage{siunitx}
\DeclareSIUnit{\gauss}{G} % Declares "G" as the symbol for Gauss
\begin{document}

\preprint{APS/123-QED}

\title{A simple magnetic field stabilization technique for atomic Bose-Einstein condensate experiments}
\author{S. Tiengo}
\author{R. Eid}
\author{M. Apfel}
\author{G. Brulin}
\author{T. Bourdel}
\thanks{thomas.bourdel@institutoptique.fr}
\affiliation{%
 Universit\'e Paris-Saclay, Institut d'Optique Graduate School,\\
 CNRS, Laboratoire Charles Fabry, 91127 Palaiseau, France 
}%

\date{\today}

\begin{abstract}
We demonstrate a simple magnetic field stabilization technique in a Bose-Einstein condensate experiment. Our technique is based on the precise measurement of the current fluctuations in the main magnetic field coils and amounts to their compensation using an auxiliary coil. It has the advantage of simplicity as compensation is done using a low inductance coil that can be straightforwardly driven at the relevant frequencies ($<$1\,kHz). The performances of the different components (power supply, current transducer, electronics...) are precisely characterized. In addition, for optimal 
stability the ambient magnetic field is also measured and compensated. The magnetic field stability around \SI{57}{G} is measured by Ramsey spectroscopy of magnetic field sensitive radio-frequency transition between two spin states of potassium 39 and the shot-to-shot fluctuations are reduced to \SI{64(7)}{\micro\gauss} rms, i.e. at the $1\times 10^{-6}$ level. In the context of our experiment, this result opens interesting prospects for the study of three-body interactions in Bose-Einstein condensate potassium spin mixtures. 
%\begin{description}
%\end{description}
\end{abstract}

\maketitle

%\tableofcontents

\section{Introduction}
A highly stable magnetic field is desirable in many contexts and in particular in quantum technologies, including quantum computing \cite{Cirac1995}, simulation, and sensing \cite{Ludlow2015}. Indeed, the coherence time of quantum systems is often reduced because of magnetic field fluctuations. In particular, neutral atoms or ions are sensitive to the magnetic field because of the Zeeman effect. Even if one works at a 'magic' condition where the Zeeman effect cancels at first order, there is typically also a contribution to second order that is relevant \cite{Sarkany2017}. As a consequence, controlling the magnetic field is crucial in any precision experiment using atoms or ions. This is especially true for clocks \cite{Ludlow2015}. 

Apart from precision measurements, ultracold atoms are also used in the context of quantum simulation, i.e. the study of well controlled many-body systems \cite{Bloch2008, Li2024}. The control of the interatomic interaction through magnetically controlled Feshbach (molecular) resonances has led to several major advances in the field \cite{Chin2010}. The use of narrow resonances requires a highly stable magnetic field \cite{Marte2002, Ciamei2022, Borkowski2023}. A stable magnetic field is also crucial when working with spin mixtures, either with a magnetic field close to zero to bring the Zeeman energy scale down to the magnitude of the spin-spin interaction energy scale, \cite{Pasquiou2011, Rogora2024} or for the precise control of radio-frequency induced transitions between hyperfine levels \cite{Cominotti2024}. In the latter case, the magnetic field-induced shifts of the transition frequency should be compared to the Rabi-coupling frequency. 

The above needs have led to the development of several techniques for magnetic field stabilization. High magnetic permeability materials such as $\mu$-metals are often used as passive magnetic field shields. They are able to efficiently reduce magnetic fluctuations by several (typically $\sim$6) orders of magnitude \cite{Farolfi2019}. They are used in most metrology experiments and even in some experiments with quantum gases. However, the need for an enclosed geometry and the incompatibility of magnetic shields with strong fields due to saturation effects have driven the development of alternative active stabilization methods \cite{Dedman2007, Xiao2020}. The magnetic field can be measured directly with sensors in close proximity to the atoms, and the signal fed back to compensation coils. The magnetic field can then be either stabilized to a low value in a given direction or set to zero in all directions. Due to the lack of high precision fluxgate sensors at high magnetic fields, such a technique is limited to low magnetic fields (typically below \SI{10}{\gauss}) \cite{Xu2019,Duan2022, Hesse2021}.

Some experiments require a stronger magnetic field. It is then usually created with a current (of typically 10 to \SI{500}{\ampere}) flowing through a pair of coils in Helmholtz configuration. A primary source of magnetic fluctuations are then the current fluctuations. In fact, the best commercially available current supplies have relative current stability in the $10^{-5}$ range (also depending on the inductance of the coils). The current can be measured to a better accuracy with transducers, and feedback can be applied to the power supply command or to a current controlling MOSFET in order to reduce the current fluctuations \cite{Merkel2019, Thomas2020, Borkowski2023}. In addition, environmental magnetic noises such as the 50/60\,Hz power line noise or slower drifts can be measured and corrected \cite{Merkel2019, Borkowski2023}. The experiments are typically also triggered to the 50/60\,Hz line.

%In this paper, we demonstrate a simple technique to improve the magnetic field stability. In short, we measure the residual fluctuation of the current in the Feshbach coil and we send a current proportional to the error signal in an auxillary coil in order to commpensate the magnetic field fluctuation. It simplify the setup it relaxes the constrain on a direct feed-back loop to the current which is complex because of the inductance of the Feshbach coils and potential instabilities. 

%We measure the reduction of the magnetic field noise using Ramsey spectroscopy and achieve 100\,$mu$G shot-to-shot rms fluctuation of a magnetic field around 50\,G, corresponding to 2ppm. It shows that our simple stabilization scheme achieves  performances at the state of the art. Interestingly, it can be straightforwardly added to an existing system where the current in the Feshbach coil is already servoed to further improve the magnetic field stability. 

In our potassium 39 quantum gas experiment, we aim at studying the properties of Bose-Einstein condensates (BEC) in coherent mixtures of hyperfine states with different scattering lengths. A pair of bias coils in Helmholtz configuration produces a bias magnetic field that is used to control the interatomic interaction. In particular, between 33 and \SI{59}{\gauss}, two hyperfine states offer the specifically interesting situation where the intra-species scattering lengths are positive whereas the inter-species scattering length is negative \cite{DErrico07, Petrov15}. When a radio-frequency field is frequency swept to bring the condensate in a dressed state, i.e. a coherent superposition of the two spin states, the global mean-field interaction is reduced and higher order terms, such as three-body interactions, may dominate the BEC equation of state \cite{Lavoine2021, Hammond2022}. Precise study of the consequences on the gas dynamics requires us to work at low Rabi frequencies \si{\ohm} (of the order of some kilohertz) as the three-body terms increase with $1/\Omega$. An effective control of the detuning of the radio-frequency dressing field as compared to the Rabi frequency requires magnetic field fluctuations in the range of \SI{100}{\micro\gauss}, whereas we typically measure magnetic fluctuations of $\sim$\SI{1}{\milli\gauss}.

In order to achieve such a performance, we have developed a simple alternative technique to improve the magnetic field stability. In short, we measure the fluctuations of the current in the bias coils and a small proportional current is driven in an auxiliary coil to compensate directly on the magnetic field for these fluctuations. In contrast to a direct proportional-integral-derivative (PID) regulation on the coil current, the proportional gain of the compensation current needs to be precisely set. Nevertheless, our method greatly simplifies the magnetic field stabilization system, as a direct feedback on the current typically has a complex frequency response in the noise frequency range \cite{Merkel2019}. In our system, we quantify the reduction of the magnetic field noise with the atoms using Ramsey spectroscopy on a radio-frequency transition between two spin states. With the addition of a compensation of the ambient magnetic field noise, we achieve \SI{64(7)}{\micro\gauss} shot-to-shot rms fluctuations of the magnetic field, corresponding to $\sim$1\,ppm. %Our technique can be straightforwardly implemented on an existing system where the current in the main coils is already servo-controlled to further improve the magnetic field stability. 
Finally, we also analyze the residual magnetic field noise over hours by performing repetitive measurements of the full Ramsey oscillations as a function of time. 
It confirms a ppm stabilization of the magnetic field on such a time scale and permits a more detailed characterization of the residual noise in our setup.

\section{Current fluctuation compensation setup}

\begin{figure}
\includegraphics[width=\columnwidth]{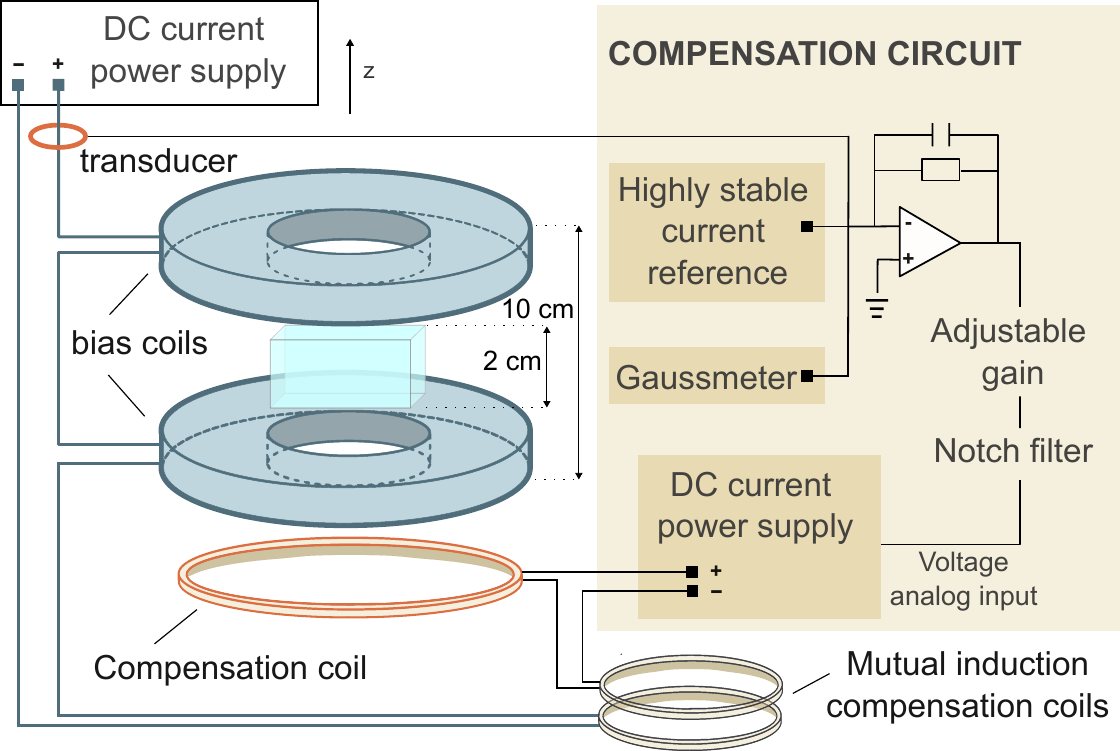}% Here is how to import EPS art
\caption{\label{fig:FeedCircuit} Schematics of the magnetic field noise compensation circuit. Ultracold $^ {39}$K atoms are prepared approximately at the center of a glass cell. Two bias coils in Helmholtz configuration generate a magnetic field set in the range 33 to \SI{59}{\gauss} at the atomic cloud position. A typical value of the current delivered in the bias coils by the stable DC power supply is of \SI{15}{\ampere}, with relative current stability at the $1\times 10^{-5}$ level. A current transducer monitors the main power supply current and its signal is compared with a highly stable DC current reference. The difference gives an error signal that is used as an input for an auxiliary current supply that drives current through a compensation coil, thereby stabilizing the magnetic field. In addition, a gaussmeter is placed close to the atoms, and its signal is added as a supplemental input to the circuit in order to also compensate ambient magnetic field fluctuations. In order to null the mutual induction between the bias coil and the compensation coil circuits, another couple of two small coils is added far from the atom position. %Each of these coils is connected in series respectively with the bias coils and the compensation coil and they are positioned far from the vacuum cell. The number of loops of this new couple of coils is adjusted so that the mutual induction coupling between them matches (with opposite sign) the coupling between the bias coils and the compensation coil.
}
\end{figure}
A diagram of the noise stabilization circuit is shown in Fig. \ref{fig:FeedCircuit}. Ultracold $^{39}$K atoms are prepared in the vacuum glass cell around which two water-cooled bias coils in Helmholtz configuration, with a total DC resistance of \SI{80}{\milli\ohm} and an inductance of $\sim$\SI{300}{\micro\henry}, produce a bias magnetic field that is used to tune atom interactions. The current through the coils is provided by a low-noise DC current power supply \textit{(HighFinesse UCS 20A/10V with the ultra-high stability option)} with a maximum current of \SI{20}{\ampere} corresponding to \SI{76}{\gauss}. To monitor the current in the coils, we use a current transducer \textit{(LEM ultrastab IT 200-S)}. A sensitivity of \SI{3}{\milli\ampere}$/$A is reached by winding the cable 3 times around the sensor. This sensor has very good noise characteristics (significantly below its specifications of 2\,ppm of full scale between 0 and 1\,kHz, see below for our measured value), although it exhibits clock noise at \SI{17}{\kilo\hertz} and harmonics. The noise spectrum of the bias coils current is found to be broad below \SI{1}{\kilo\hertz}, and with its main frequency components around \SI{300}{\hertz} \footnote{Note that we also have added a \SI{1}{\micro\farad} capacitor in parallel to the power supply to slightly reduce the current noise and also its typical frequencies}. To quantify the DC current noise we use a 6.5 digits multimeter (\textit{Keysight 34465A}) with \SI{176}{\us} integration time, measuring the rms of the signal after the current transducer. This technique has the advantage of being insensitive to \SI{17}{\kilo\hertz} clock noise, while being sensitive to the current noise in the relevant frequency range. At a current of \SI{19}{\ampere}, from repetitive measurements during a few seconds, we measure a relative rms noise of $\sim10^{-5}$. Note that this already quite good performance\footnote{A small fraction of this relative noise ($\sim$3\,$10^{-6}$ rms) originates from the power supply's external analog input which is driven by one output channel of a digital-to-analog converter card (\textit{National Instrument PCIe-6738})} is directly obtained from a commercial power supply which, in fact, already includes a proportional-integrator (PI) feedback loop to stabilize its current \footnote{In this configuration, trying to further stabilize the current by applying some feedback to its analog command quickly leads to current oscillations.} . On longer time scales (hours), the mean value of the current also exhibits additional offset jumps at a few $10^{-5}$ level.

We first focus our magnetic field stabilization strategy on the precise measurement of the current fluctuations and on their compensation. Compensation is performed by sending a current proportional to the measured noise in a compensation coil. The input signal for the compensation circuit is a current error signal, obtained from the subtraction of a highly stable DC current reference from the current provided by the above-mentioned current transducer. The reference current is produced by a precision voltage source (\textit{Stanford Research system DC205}, stability 1\,ppm/24h, 0.3\,ppm noise) connected to a highly stable 100\,$\Omega$ resistance (\textit{VPG Foil Resistors Y0926100R000T9L}, $\pm$2\,ppm/K). The difference signal serves as an input for an inverting transimpedance amplifier circuit (OP277BA) with a \SI{1}{\kilo\ohm} feedback resistor. A \SI{33}{\nano\farad} capacitor is added in parallel to the resistor to limit the current noise detection bandwidth, low-pass filtering the signal with a \SI{4.8}{\kilo\hertz} cut-off frequency. Note that our method has the advantage that the transducer current is directly subtracted from a reference signal (before any amplification), therefore avoiding the addition of electronic noise prior to the generation of the error signal. A notch filter at \SI{17}{\kilo\hertz} further suppresses the transducer clock noise and a variable gain amplifier enables the adjustment of the overall gain for the compensation circuit. The measured noise originating from the different elements are listed in table\,\ref{tabular:NoiseTable}, showing that we can indeed precisely measure the current fluctuations from the power supply. 
\begin{center}
\begin{table}

\begin{tabular}{|c|c|c|}
\hline
Noise origin & rms noise [\si{\micro\gauss}]) & triggered to \SI{50}{\hertz}  \\ 
\hline
Power-supply (\SI{19}{\ampere}) & 630 & 190 \\  
 \hline
Current transducer & 39 & 23 \\
\hline
Reference signal & 33 & 16 \\
\hline 
noise from electronics & 28 & 6 \\
\hline
\end{tabular}
\caption{Rms noises measured at the output voltage of the compensation circuit. The values are extracted from repetitive measurements (over few seconds) using the precision voltmeter with a \SI{176}{\us} integration time (triggered or not to the \SI{50}{\hertz} line). They are then scaled to magnetic field unit. The power-supply current noise is measured with both the transducer and reference current signals connected. The baseline noise of the current transducer is measured with the transducer serving as the single input of the circuit and the main DC current supply turned off. The reference noise is measured with the reference signal as the only input and at a low reference current input (low DC reference voltage) in order to avoid saturation of the first amplifier. The electronics noise is measured with a \SI{50}{\ohm} resistance as unique input of the circuit.}
\label{tabular:NoiseTable} 
\end{table}
\end{center}

The generated voltage error signal, ranging between 0 to \SI{10}{\volt}, is used as an external analog modulation voltage to a laser current supply (\textit{Thorlabs LDC 8005}), which can drive a current between 0 and $\pm$\,\SI{500}{\milli\ampere} through the compensation coil (calibration: 2.0\,G/A). This current supply, even connected to the compensation coil, has a \SI{20}{\kilo\hertz} bandwidth and a flat frequency response in the frequency range of interest. This is a key advantage of our method as compared to a direct feedback to the main coil circuit, which typically has a complex response in the frequency range of its noise \cite{Merkel2019}. However, we first have observed that mutual inductive coupling between the main coils and the compensation coil has significant effects and was leading to instabilities of the current compensation system as soon as it was activated. In order to solve this problem, we cancel the mutual induction using an opposite sign mutual coupling between two superimposed coils, each of them in series with either the main coils or the compensation coil (see Fig.\,1). Experimentally, the induction can be quantified by measuring the induced voltage in one circuit while modulating the current in the other. The size and number of turns in the mutual inductance compensation coils are adjusted to zero the total mutual induction between the two circuits. 

Importantly, we need the magnetic stabilization to work in the presence of radio-frequency fields (between 30 and \SI{50}{\mega\hertz}) used for spin state control. Such fields are found to cause electromagnetic interferences on electronic circuits that are difficult to shield. In practice, in addition to inducing noise at the same frequency, they also lead to small unstable DC offsets in the electronics signals. We solve this issue by moving the main current supply for the Helmholtz coils and the whole compensation circuit to a different room from the one where the vacuum system and the atoms are. 

\section{Ramsey interferometry measurement of the magnetic field}

The performance of the stabilization circuit is measured through Ramsey spectroscopy \cite{Ramsey50} of the transition between the $\ket{\uparrow}=\ket{F=1, m_F=-1}$ and $\ket{\downarrow}=\ket{F=1, m_F=0}$ hyperfine ground states in $^{39}$K atoms. The magnetic field is set to \SI{57}{\gauss} at which the energy gap between the two levels is about \SI{40}{\mega\hertz} and depends linearly on the magnetic field with \SI{0.70}{\mega\hertz}/G. We optically trap $N\approx1\times 10^6$ ultracold atoms in the $\ket{\uparrow}$ state at a temperature of \SI{2}{\micro\kelvin}. The harmonic trap frequencies are $(62\times300\times 300)$\,Hz. The radio-frequency field is created by a simple two-loop antenna with a diameter of \SI{2}{\cm}, placed $\sim$\SI{3}{\cm} away from the atoms. Ramsey sequences are performed using radio-frequency $\pi/2$ pulses with a Rabi frequency of \SI{10}{\kilo\hertz} (pulse time $\tau=$\SI{25}{\us}). Finally, the optical trap is turned off and another set of coils in quadrupole configuration produces a strong magnetic field gradient, maintained for \SI{2}{\ms} to separate the two spin states by the Stern-Gerlach effect. All currents are then switched off, and, after \SI{1.3}{\ms}, the magnetic field is low enough to image the atoms by resonant fluorescence imaging. In each repetition of the experiment, we can measure the fraction $f$ of atoms that remain in the initial $\ket{\uparrow}$ state. Note that this measurement is not sensitive to total atom number fluctuations and with our signal to noise we estimated a precision of 0.2$\%$, as can be measured using a single $\pi/2$ pulse producing a stable 50/50 mixture \footnote{The $\pi/2$ pulse actually puts each atom in a coherent spin superposition. It theoretically results in binomial distribution of the number of atoms in each spin state with rms fluctuations of $f$ that amount to $1/2\sqrt{N}=$0.05\%. Our measured detection noise is only 4 times this shot-noise limit.}. 

In practice, we slightly detune the radio-frequency from atomic resonance such that we detect oscillations of the atomic fraction as a function of the Ramsey time $T$: 
\begin{equation}\label{single_shot}
\begin{split}
f &\approx \frac{1}{2} \left(1- C(T)  \cos\left(\delta T\right)\right)\\
\text{with\;} \,T &= T_{\text{wait}} + 4\tau/\pi,
\end{split}
\end{equation}
where $\delta$ is the radio-frequency detuning from resonance and $C(T)$ the contrast of the interferometer, that might depend on $T$. The above formula \cite{Vanier} assumes that the detuning is much lower than the Rabi frequency and includes a correction on the Ramsey time $T$ due to the finite pulse time $\tau$. 
The highest magnetic field sensitivity of the signal $f$ is achieved at half fringe, for example when $\delta T \approx \pi/2$. For small deviations around that point, $f$ is linear with the magnetic field noise $f - 1/2 \simeq C(T)(\delta T-\pi/2)/2$. This expression for $f$ also assumes that $\delta$, or equivalently the magnetic field is constant throughout the Ramsey sequence, which is valid as long as $T$ is short as compared to the inverse fluctuation frequencies of the B field. For $T_{\text{wait}}$=\SI{100}{\us} and $\delta \approx 2$\,kHz the first half fringe position of $f$ is experimentally realized and here we take repetitive shots to measure to the shot-to-shot variation of the magnetic field. After \SI{100}{\micro\second} of waiting time, the contrast $C(T)$ of the Ramsey fringes is 0.95(1) (see Sect. \ref{RamseyAnalysis}). Our 0.2\,\% precision on $f$ corresponds to a single-shot precision of \SI{5.3}{\hertz} on the detuning $\delta/2\pi$, or equivalently, of \SI{7.6}{\micro\gauss} at each repetition of the experiment. Without compensation of the magnetic field noise, we measure magnetic field rms fluctuations of \SI{0.8}{\milli\gauss} over 30 repetitions of the experiments. A single experimental sequence approximately takes 11 seconds to complete, hence the acquisition of 30 shots requires about 5 minutes. In addition, we confirm that the magnetic field noise measured by atomic Ramsey spectroscopy is well correlated with a simultaneous measurement of the bias coil's current. 

We now turn to the measurement of the magnetic field with our circuit activated, compensating for the noise of the bias coil current. For the optimal compensation of the magnetic field noise, the variable gain for the compensation current has to be precisely set. This is done by finding the gain for which the magnetic field measured with the atoms is insensitive to a small intentional variable offset in the current of the bias coils. We estimate that the gain is set with a precision of about 1\,\%. With compensation of the current noise, we then find a reduced rms noise (over 30 repetitions of the experiment and over 5 minutes) corresponding to \SI{136(18)}{\hertz} in frequency or equivalently \SI{195(26)}{\micro\gauss} in magnetic field.

\section{External magnetic field noise and its compensation}
In order to analyze the additional variations of the external magnetic field in the vertical direction, a gaussmeter (\textit{Bartington Mag-03}, bandwidth \SI{3}{\kilo\hertz}) is placed in the proximity of the atoms. It first detects \SI{50}{\hertz} and harmonics originating from transformer coils connected to the power line. Actually, already for the previous measurements, power supplies fed by the \SI{50}{\hertz} power line were positioned away ($\gtrsim$\,\SI{2}{\m}) and oriented in such a way to minimize their contribution to the vertical magnetic field at the atom position. This noise contribution can alternatively be zeroed by synchronizing the experiment to the 50\,Hz line. In addition, low-frequency drifts (below \SI{0.5}{\hertz}) of the external magnetic field field are identified as the main limitation to our stability.  We finally choose the precise gaussmeter position about \SI{50}{\cm} away from the atomic position such that the measured magnetic field is small (below 1\,G) and  only marginally affected by the current in the bias and compensation coils in order to disentangle the effects from the current noise and from the ambient magnetic field.

We then add the measured external magnetic field to our compensation setup. The gaussmeter output voltage is connected through a \SI{2.5}{\kilo\ohm} resistance (a value chosen to have the proper gain in the compensation) as a supplementary input to the transimpedance amplifier. It now has 3 inputs: the current from the current transducer, the current from the voltage reference and the current originating from the gaussmeter. We then repeat the Ramsey measurements  (over 30 shots and 5 minutes), now also synchronized to the \SI{50}{\hertz} line, and found a residual fluctuation of the resonance transition frequency of \SI{45(5)}{\hertz} rms, corresponding to \SI{64(7)}{\micro\gauss} rms. This value is only a factor $\sim$\,2 higher than the noise floor of the compensation system (\SI{28}{\micro\gauss} if we add the noise contributions from the transducer and from the reference, see TABLE \ref{tabular:NoiseTable}). 
Note that our compensation is not really efficient in reducing the \SI{50}{\hertz} modulations, as they differ at the gaussmeter and at the atom position. This is not necessarily a problem since these modulations are first small (if one is careful with the power supply positions) and second deterministic once the experiment is synchronized to the \SI{50}{\hertz} line. If necessary, they could be precisely measured with Ramsey spectroscopy and then feed forward to the compensation circuit \cite{Merkel2019}.

\section{Analysis of Ramsey oscillations on hour time scale}\label{RamseyAnalysis}
\begin{figure}
\includegraphics[width=\columnwidth]{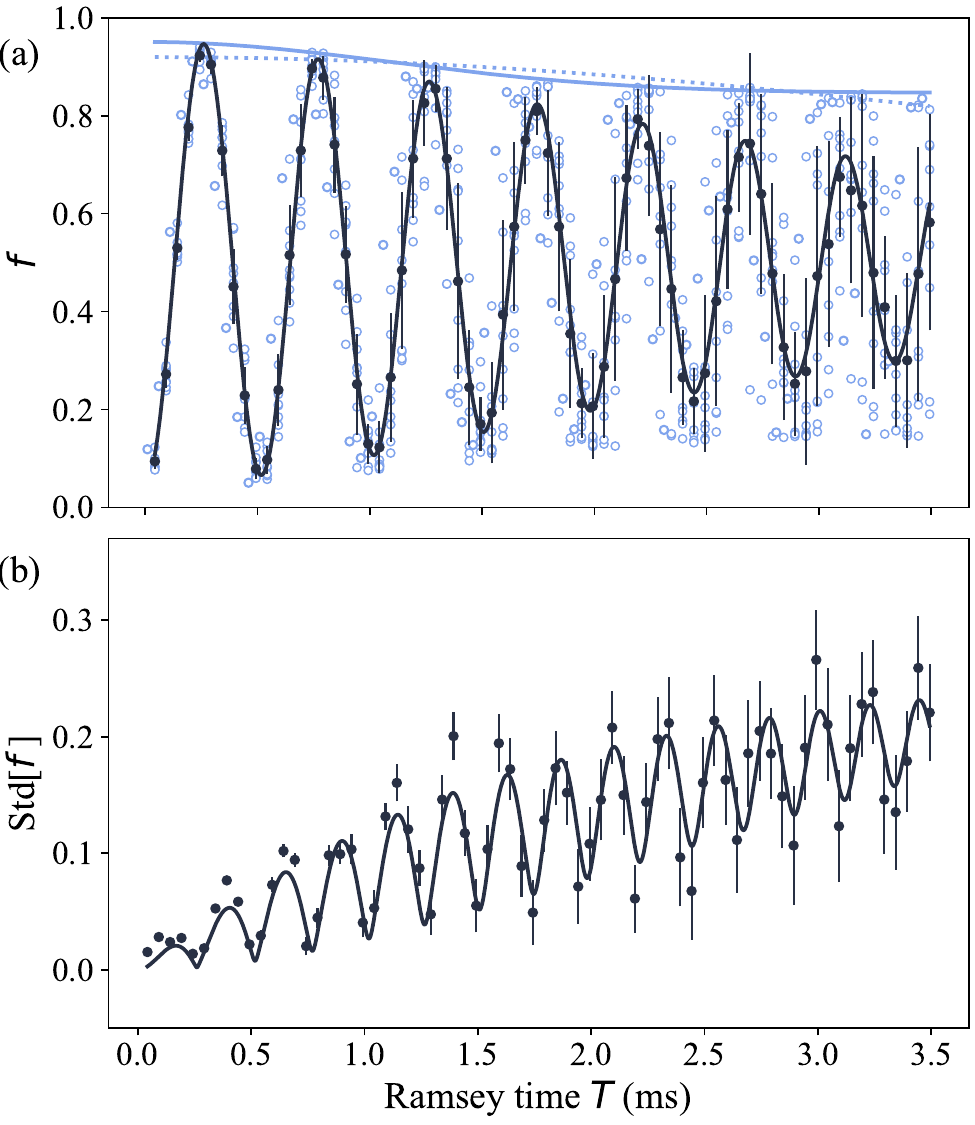}
\caption{\label{fig:Ramsey_DV} 
Repetitive measurements of Ramsey oscillations as a function of time $T$ up to 3.5 ms. (a) Individual measurements of the transfer fraction $f$ (blue circles). The envelope of the oscillation signals allow us to extract the contrast $C(T)$ of the interferometer which is found to be better fitted by a Gaussian with an offset (solid blue line) rather than a simple Gaussian decay (dashed blue line). The average of the 10 measurements $\textrm{Mean}[f]$ with their standard deviations $\textrm{Std}[f]$ is also plotted (black dots) and is fitted (solid black line) in order to extract the magnetic field fluctuations (see text). (b) Standard deviation of the atom fraction $f$ as a function of the waiting time $T$ from 10 repetitions of the Ramsey experiment. The error bars are estimated from simulated data using Monte-Carlo sampling of the noise because it is difficult to extract them from only 10 experimental realizations. The fit with equation\,\ref{eq:var} permits us to evaluate the magnetic fluctuations.}
\end{figure}

We now turn to the measurement of the magnetic field stability over hours. Rather than repeating the previous Ramsey spectroscopy measurements, we chose to perform repetitive measurements of Ramsey oscillations of the atomic fraction as a function of time and up to $T=$\,\SI{3.5}{\ms} (see Fig. \ref{fig:Ramsey_DV}), as it gives complementary information and brings additional ways of assessing the magnetic field stability. Each Ramsey sequence counts 70 data points (see Fig.\,\ref{fig:Ramsey_DV}(a)). Given the typical duration of 11 seconds for each experimental sequence, the acquisition of 10 full datasets with a total of 700 data points requires 2 hours. 

Looking at all the transferred fractions $f$ (blue circles in Fig.\,\ref{fig:Ramsey_DV}(a)), it is clear that they lie in an envelope function that we interpret as the single shot contrast $C(T)$ of the interferometer (solid blue line). Indeed, according to the formula (\ref{single_shot}), single points lies between $(1-C(T))/2$ and $(1+C(T))/2$. 
The reduction of the contrast $C(T)$ over time does not come from shot-to-shot magnetic field fluctuations but rather from an inhomogeneous broadening of the atomic resonance or from decoherence effects. For instance, a residual magnetic field gradient leads to an inhomogeneous broadening of the detuning $\delta$ because of the Gaussian spatial atomic distribution. We have indeed observed that the decay of $C(T)$ has been significantly reduced after nulling a residual magnetic field gradient (initially of \SI{8.9}{\gauss}/m) along the elongated direction of the cloud. The remaining decay could be due to the magnetic field gradients in the other two directions that we did not compensate and/or to a mean-field shift that depends on the density (we estimate the mean-field shift of the resonant frequency to be \SI{150}{\hertz} at our peak density). In addition, we find that a simple Gaussian decay (dashed blue line) does not perfectly fit the measured contrast decay, which appears almost constant in the range \SI{2}{ms}\,$<T<$\,\SI{3.5}{\ms}. This behavior can probably be associated with the atom radial oscillations in the optical trap leading to a partial time averaging of the inhomogeneous frequency shifts in these directions at long Ramsey times $T$ only. 

The shot-to-shot variation of the magnetic fields is reflected in the dephasing of the Ramsey oscillations. After averaging over the 10 repetitions of the experiment (black dots in Fig.\,\ref{fig:Ramsey_DV}(a)), this leads to a further decay of the contrast. 
Assuming a Gaussian distribution of the radio-frequency detuning with mean $\delta_0(T)$ and standard deviation $\sigma$, we fit the averaged data with the following function
\begin{equation}
\textrm{Mean}[f] = \frac{1}{2} \left(1- C(T) \cdot e^{-\frac{1}{2}(\sigma T)^2} \cdot \cos\left(\delta_0(T) T+ \phi \right)\right).\label{eq:mean}
\end{equation}
We include the possibility of a slight variation of $\delta_0(T)$ as a function of $T$ as it is visible after careful data analysis. In fact, it comes from deterministic \SI{50}{\hertz} ambient magnetic field fluctuations. \footnote{Precisely, $\delta_0(T)$ is the ensemble and time averaged of the Ramsey detuning over the time $T$.} In practice, since the maximum Ramsey time 3.5\,ms is small as compared to 20\,ms, we can assume a linear behavior $\delta_0(T)=\delta_{0}(T_{\text{wait}}=0)+\alpha T$. From the fit, we extract $\delta_{0}(T_{\text{wait}}=0)=2\pi\times$\SI{1900(5)}{\hertz}, $\alpha=2\pi\times$\SI{59(3)}{\hertz}/ms and the standard deviation of the detuning $\sigma$, corresponding to a residual magnetic field rms noise \SI{71(4)}{\micro\gauss}. This value is only slightly higher than the one obtained on 30 shots and 5 minutes, showing that we are also able to maintain excellent magnetic field stability over two hours. Note that our measured magnetic field  stability also implies an excellent mechanical and thermal stability of the bias coils, as it corresponds to a variation of distance between the two coils of $\sim$\SI{0.1}{\um} \footnote{In our experiment, we found that the good long term stability could only be observed with at least \SI{2}{\s} wait time between the magnetic quadrupole trap step, where the bias coils are used with \SI{200}{A} for about \SI{1}{\s}, and the Ramsey interferometer sequence. This observation could have both a mechanical (because of the force between the two coils) or a thermal (because of the heat deposited in the coils) origin.}. 

We conclude our analysis of the Ramsey fringes data with the study of the standard deviations of $f$ as a function of time $T$ (see Fig. \ref{fig:Ramsey_DV}(b)). For Gaussian noise, the expected function describing the standard deviation of $f$, can be derived 
\begin{align}
%\frac{\textrm{Var}[f]}{C(T)^2}= 
\frac{\textrm{Std}[f]^2}{C(T)^2} =&\frac{1}{8}\left( 1+e^{-2\sigma^2T^2}\cos (2\delta_0(T) T+2\phi)\right) \notag \\
& -\frac{1}{4}e^{-\sigma^2T^2}\cos^2(\delta_0(T) T + \phi)\label{eq:var}
\end{align}
We then fit this function to our measured standard deviation and extract a residual noise corresponding to \SI{69(2)}{\micro\gauss}, a value that is in agreement with the previous estimation. We can note a tendency of the fit to underestimate the measured standard deviation at short times $T$. This may indicate that the amount of magnetic field noise slightly depends on the integration time $T$ as can be expected if there is residual noise in the frequency range \SI{300}{\hertz}-\SI{1}{\kilo\hertz}, that is averaged out at long Ramsey time $T>1$\,ms. The overall quality of the fit assuming a constant $\sigma$ nevertheless indicates that the residual magnetic field noise is dominated by shot-to-shot fluctuations. 

%Interestingly, although we did not precisely monitor the magnetic field fluctuations for weeks, the magnetic field stability seems to be preserved over such time scale as we continue to work with the compensation setup.

\section{Conclusions}
In conclusion, we have shown a new method for stabilizing the magnetic field around \SI{57}{\gauss} without the use of a magnetic field shield. It mainly amounts to a compensation the fluctuations of current in the coils that produce the magnetic field. This is done through its precise measurement and driving a compensation current in a low inductance auxiliary coil. Such a method has the advantage to circumvent the problem of the complex frequency response of the main circuit due to the high inductance of the main coils. To further improve the magnetic field stability, we also have implemented a compensation of the ambient magnetic field noise. Detailed characterization of our system performances through repetitive Ramsey measurements of the magnetic field using the atoms was performed and we find rms magnetic field fluctuations of \SI{64(7)}{\micro\gauss} on 5 minutes and of \SI{71(4)}{\micro\gauss} on two hours.
Although we have not taken specific data to assess it, given the specifications of the circuit components, good stability is expected and indeed experimentally seems to be maintained over days and weeks. Interestingly, our method can be straightforwardly applied to other systems in order to improve the magnetic field stability. In the context of our experiment, such a magnetic field stability opens prospects for the study of Bose-Einstein condensates in a coherent spin superposition because of dressing by a radio-frequency field. In particular, the consequences of the emerging three-body interactions \cite{Lavoine2021, Hammond2022} in the equation of state of condensates will be studied.

\begin{acknowledgments}
We thank L. Lavoine and A. Hammond for early developments on current noise measurements. This  research  was  supported  by  CNRS,  Minist\`ere  de  l'Enseignement  Sup\'erieur  et  de  la  Recherche, Labex PALM, Quantum Paris-Saclay, R\'egion Ile-de-France  in  the  framework  of  Domaine d'Int\'er\^et Majeur Quantip, PEPR Quantique Dyn1D, Paris-Saclay in the framework of IQUPS, ANR Droplets (Grant No. 19-CE30-0003), and the Simons Foundation (Award No. 563916,  localization of waves).
\end{acknowledgments}

%\appendix

%\section{Appendixes}
%\section{A little more on appendixes}
%\subsection{\label{app:subsec}A subsection in an appendix}

\bibliography{references2}% Produces the bibliography via BibTeX.

\end{document}